\documentclass[aps,prb,showpacs,twocolumn,floats]{revtex4}
\usepackage{amssymb}

\usepackage{amsmath}
\usepackage{graphicx,psfig}
\usepackage{dcolumn}
\usepackage{bm}

\input{tcilatex}

\begin{document}

\title{Dipolar ordering in crystals of Mn$_{12}$ Ac}
\author{D. A. Garanin}
\affiliation{ \mbox{Department of Physics and Astronomy, Lehman College, City
University of New York,} \\ \mbox{250 Bedford Park Boulevard
West, Bronx, New York 10468-1589, U.S.A.} }
\date{\today}

\begin{abstract}
Ordering in realistic elongated box-shape crystals of the molecular magnet Mn%
$_{12}$ Ac is investigated with the site-resolved mean-field approximation
that does not assume a uniform ordering. It is shown that ferromagnetic
ordering should not occur in crystals with the aspect ratio up to
12. Instead, for the aspect ratio about 6, the inner and outer regions of
the crystal order in different directions, uniformly along its length.
Finding ordering temperature by extrapolating the inverse susceptibility
curve does not provide a correct $T_{C}.$
\end{abstract}
\maketitle


\bigskip

Magnetic ordering in molecular magnets such as Mn$_{12}$ Ac,\cite{lis80}
attracts attention of researchers because the ordering dynamics, as well as
domain-wall motion,\cite{garchu08prb} might be intimately related to the
spectacular phenomenon of resonance spin tunneling\cite{frisartejzio96prl}
under the energy barrier created by the strong uniaxial anisotropy\cite
{sesgatcannov93nat} acting on the effective spin $S=10$ of the magnetic
molecule. The anisotropy barrier is responsible for the bistability of
magnetic molecules, so that at temperatures below 10 K the latter can be
considered as Ising spins 1/2 pointing up or down. Mn$_{12}$ Ac crystallizes
into a body-centered tetragonal lattice with the periods $a=b=17.319$ \AA\
and $c=12.388$ \AA , $c$ being the direction of the magnetic easy axis $z$.
The unit-cell volume is $v_{0}=abc=3716$ \AA $^{3}$ and there are two
molecules per unit cell, $\nu =2$.

Since the magnetic core of magnetic molecules is surrounded by organic
lygands, there is practically no exchange interaction between the molecules.
Thus the sole source of magnetic ordering is the dipole-dipole interaction
(DDI) that leads to ordering below 1 K in Fe$_{8}$ (Refs.\
\onlinecite
{feralo00prb,marchu00jpcm}), Mn$_{12}$ Ac (Ref.\ \onlinecite{luietal05prl}),
and other compounds. \cite
{evaetal04prl,evaetal06prl,moretal06prb,belborpow06prb} The type of dipolar
magnetic ordering in molecular magnets is a subtle question. Since a column
of spins up directed along the $c$ axis produces a magnetic field up on its
own spins that largely exceeds the field it produces on other spins, \cite
{feralo00prb} one concludes that the ground state includes ferromagnetically
ordered columns along the $c$ axis. The ferromagnetic columns can order
ferro- or antiferromagnetically with respect to each other.

It was shown that the ground state of ellipsoids of revolution depends on
their aspect ratio. \cite{garchu08prb} For ellipsoids very prolate along the
$c$ axis, the ground state was found to be ferromagnetic, whereas in other
cases there are ferromagnetic columns or planes ordered
antiferromagnetically with respect to each other. The energies of
differently ordered states are rather close, so that practically the system
will order in a spin-glass state upon fast cooling.

Evidence for a ferromagnetic ordering in Mn$_{12}$ Ac was obtained in Ref.\ %
\onlinecite{luietal05prl} by neutron scattering experiments. However, the
shape of the crystal was not specified. Usually crystals of molecular
magnets are box-shaped and elongated along the $c$ axis but it is a question
if they are elongated enough to make ferromagnetic ordering prevail. Recent
measurements of the ferromagnetic susceptibility $\chi $ above $T_{c}$ in Mn$%
_{12}$ Ac and its modifications\cite{bowetal09-arXiv} were done on box-shape
single crystals with the aspect ratio $L_{z}/L_{x}$ of about 5-6. In the
absense of transverse field, one cannot approach $T_{c}$ because the
relaxation time becomes too long below the blocking temperature 3 K due to
the anisotropy barrier. In this case $T_{c}\simeq 0.9$ K was obtained by the
linear extrapolation of the $\chi ^{-1}(T)$ curve.

Applying a strong transverse field $B_{\bot }$ up to 6 T (Refs.\
\onlinecite
{luietal05prl,bowetal09-arXiv}) increases the relaxation rate via barrier
lowering and spin tunneling, so that ordering can be achieved during a
realistic time\cite{luietal05prl} and the susceptibility measurements can be
extended to lower temperatures.\cite{bowetal09-arXiv} However, the
transverse field tends to suppress magnetic ordering and lower $T_{c}$ via
the three main effects: (i) spin canting that reduces the effective magnetic
moment along the $z$ axis, (ii) tunneling hybridization of the $\left| \pm
S\right\rangle $ states that acts as the transverse field in the Ising model
that lowers $T_{c}$ and leads to quantum criticality, and (iii) random
longitudinal fields resulting from the strong transverse field and random
tilts of the molecular easy axes, further suppressing ordering. These
effects have been recently discussed\cite
{milkensaryes10prb,gar09dipolarorderingunpub} in connection with experiments
\cite{bowetal09-arXiv} on the basis of the mean-field approximation (MFA)
that should work well for long-range interactions such as DDI.

The aim of this work is to investigate the possibility of ferromagnetic
ordering in its competition with other ordering types in crystals of Mn$_{12}
$ Ac of the realistic box shape. While calculations of dipolar fields in
crystals of molecular magnets were done for ellipsoids of revolution\cite
{marchu00jpcm,garchu08prb,milkensaryes10prb} that are impossible to grow, no
theoretical work has been done yet on box-shape crystals. It is not obvious
that long rods behave similarly to long ellipsoids. It was shown that the
dipolar field in the middle of the end faces of a uniformly magnetized
cylinder of Mn$_{12}$ Ac is opposite to the magnetization (the top of left
column of page 4 of Ref.\ \onlinecite{garchu08prb}). This should lead to
spin flips at the ends of the cylinder with a subsequent proliferation into
its body. The same can be expected for elongated boxes. Since the dipolar
field in crystals of other than ellipsoidal shape is non-uniform, the MFA
equations take the form of a large system of equations for all magnetic
molecules considered separately.

The model includes pseudospin variables $\sigma _{i}=\pm 1$ for molecules at
each lattice site $i$ of a boby-centered tetragonal lattice. The magnetic
moment of a molecule is $Sg\mu _{B}$ with $g=2.$ The dipolar field on
magnetic molecule $i$ is the sum over positions of all other molecules $j$%
\begin{equation}
B_{i,z}^{(D)}=\frac{Sg\mu _{B}}{v_{0}}D_{i,zz},\qquad D_{i,zz}\equiv
\sum_{j}\phi _{ij}\sigma _{jz}.  \label{BviaD}
\end{equation}
Here $D_{zz}$ is the reduced dipolar field and
\begin{equation}
\phi _{ij}=v_{0}\frac{3\left( \mathbf{e}_{z}\cdot \mathbf{n}_{ij}\right)
^{2}-1}{r_{ij}^{3}},\qquad \mathbf{n}_{ij}\equiv \frac{\mathbf{r}_{ij}}{%
r_{ij}}.  \label{psiijDef}
\end{equation}
Inside a uniformly magnetized ellipsoid, $\sigma _{z}=\mathrm{const},$ the
dipolar field is uniform and one has $D_{zz}=\bar{D}_{zz}\sigma _{z},$ where
\begin{equation}
\bar{D}_{zz}=\bar{D}_{zz}^{(\mathrm{sph})}+4\pi \nu \left(
1/3-n^{(z)}\right) ,  \label{Dzzbar}
\end{equation}
$\nu $ is the number of magnetic molecules per unit cell ($\nu =2$ for Mn$%
_{12}$ Ac) and $n^{(z)}=0,$ $1/3,$ and 1 for a cylinder, sphere, and disc,
respectively. The reduced dipolar field in a sphere $\bar{D}_{zz}^{(\mathrm{%
sph})}$ depends on the lattice structure. For Mn$_{12}$ Ac direct lattice
summation yields $\bar{D}_{zz}^{(\mathrm{sph})}=2.155$ that results in $\bar{%
D}_{zz}^{(\mathrm{cyl})}=10.53$ for a cylinder.\cite{garchu08prb} Then Eq.\ (%
\ref{BviaD}) yields the dipolar field $B_{z}^{(D)}\simeq 52.6$ mT in an
elongated sample that was also obtained experimentally.\cite{mchughetal09prb}
The ground-state energy in the above uniform states is given by
\begin{equation}
E_{0}=-(1/2)\bar{D}_{zz}E_{D},\qquad E_{D}\equiv \left( Sg\mu _{B}\right)
^{2}/v_{0},  \label{E0Def}
\end{equation}
where $E_{D}$ is the dipolar energy, $E_{D}/k_{B}=0.0671$ K for Mn$_{12}$
Ac. The mean-field Curie temperature is given by\cite{garchu08prb}
\begin{equation}
T_{C}=E_{D}\bar{D}_{zz}/k_{B}  \label{TCsmallDelta}
\end{equation}
that results in $T_{C}=0.707$ K.

States with ferromagnetically ordered planes alternating in the $a$ or $b$
directions in \emph{each} sublattice of Mn$_{12}$ Ac have $\bar{D}%
_{zz}=9.480,$ independently of the sample shape.\cite{garchu08prb} The state
with alternating chains in each sublattice, directed along the $c$ direction
has a very close value $\bar{D}_{zz}=9.475.$ For the two-sublattice
antiferromagnetic ordering one obtains $\bar{D}_{zz}=8.102.$ Thus, in a
strongly prolate ellipsoid of Mn$_{12}$ Ac ferromagnetic ordering is
preferred. It is interesting to estimate how strongly prolate the ellipsoid
has to be for this to be the case. Equating $\bar{D}_{zz}$ of Eq.\ (\ref
{Dzzbar}) to $\bar{D}_{zz}=9.480$ for the alternating-planes structure, one
obtains the maximal demagnetizing factor $n^{(z)}=0.0419.$ Using the formula
for prolate ellipsoids of revolution, one obtains that the minimal shape
aspect ratio $R_{z}/R_{x}=6.13$ is required for ferromagnetic ordering.

The longest crystal used in the experiments of Ref.\ %
\onlinecite{bowetal09-arXiv} has dimensions $0.4\times 0.4\times 2.4$ mm$^{3}
$ and thus the aspect ratio 6 that would be still slightly insufficient for
a crystal of ellipsoidal shape to order ferromagnetically. We will see below
that even much longer box-shape Mn$_{12}$ Ac crystals \emph{do not order
ferromagnetically}. The reason for this is the above mentioned instability
of the ferromagnetic ordering at the ends driven by the negative value of
the dipolar field, $\bar{D}_{zz}=-2.03.$ \cite{garchu08prb}

The system of Curie-Weiss equations for a crystal of molecular magnet in a
uniform external field $B_{z}$ has the form
\begin{eqnarray}
\left\langle \sigma _{iz}\right\rangle &=&\tanh \frac{Sg\mu _{B}\left(
B_{i,z}^{(D)}+B_{z}\right) }{k_{B}T}  \notag \\
&=&\tanh \left( \frac{E_{D}/k_{B}}{T}\sum_{j}\phi _{ij}\left\langle \sigma
_{jz}\right\rangle +\frac{h_{z}}{T}\right) ,  \label{CWEq}
\end{eqnarray}
where $h_{z}\equiv Sg\mu _{B}B_{z}/k_{B}$. The linearized Curie-Weiss
equations above $T_{c}$ can be cast into the matrix form
\begin{equation}
\left( T\mathbb{I-V}\right) \cdot \left\langle \mathbf{\sigma }%
_{z}\right\rangle =h_{z}\mathbf{I.}  \label{SusEqMatr}
\end{equation}
Here $\mathbb{V}$ is the DDI matrix, $\left\{ \mathbb{V}\right\}
_{ij}=\left( E_{D}/k_{B}\right) \phi _{ij},$ $\mathbb{I}$ is a unit matrix, $%
\left\{ \mathbb{I}\right\} _{ij}=\delta _{ij},$ and $\mathbf{I}$ is a unit
vector, $\left\{ \mathbf{I}\right\} _{i}=1.$ In fact, position of a molecule
in the lattice is defined by three numbers $i_{a},$ $i_{b},$ and $i_{c}$
corresponding to the three directions $a,$ $b,$ and $c,$ plus the sublattice
index. The indices $i,j$ are compound indices running from 1 to the number
of sites in the lattice $N.$ In terms of $i,j$ the matrix $\mathbb{V}$ is
non-symmetric, so that care should be taken by distinguishing between its
right and left eigenvectors. One can seek the solution in the form $%
\left\langle \mathbf{\sigma }_{z}\right\rangle =\sum_{\mu }C_{\mu }\mathbf{A}%
_{\mu }^{R},$ where $\mathbf{A}_{\mu }^{R}$ are right eigenvectors, $\mathbb{%
V}\cdot \mathbf{A}_{\mu }^{R}=T_{\mu }\mathbf{A}_{\mu }^{R},$ and $T_{\mu }$
are eigenvalues, $\mu =1,\ldots ,N.$ The right and left eigenvectors satisfy
the orthonormality condition $\mathbf{A}_{\mu ^{\prime }}^{L}\cdot \mathbf{A}%
_{\mu }^{R}=\delta _{\mu ^{\prime }\mu },$ i.e., the matrix of left
eigenvectors is the inverse of that of the right eigenvectors. Using the
orthonormality, from Eq.\ (\ref{SusEqMatr}) one obtains $C_{\mu }=\left(
\mathbf{A}_{\mu }^{L}\cdot \mathbf{I}\right) h/\left( T-T_{\mu }\right) .$
For the susceptibility of the whole crystal $\chi =\left( 1/N\right) \left(
\left\langle \mathbf{\sigma }_{z}\right\rangle \cdot \mathbf{I}\right) /h$
one finally obtains
\begin{equation}
\chi =\frac{1}{N}\sum_{\mu =1}^{N}\frac{\left( \mathbf{A}_{\mu }^{L}\cdot
\mathbf{I}\right) \left( \mathbf{A}_{\mu }^{R}\cdot \mathbf{I}\right) }{%
T-T_{\mu }}.  \label{chiRes}
\end{equation}
The ordering temperature $T_{C}$ can be identified with the maximal
eigenvalue $T_{\mu },$ this is the first time as Eq.\ (\ref{chiRes})
produces infinity as temperature is lowered.

\begin{figure}[t]
\unitlength1cm
\begin{picture}(11,5)
\psfig{file=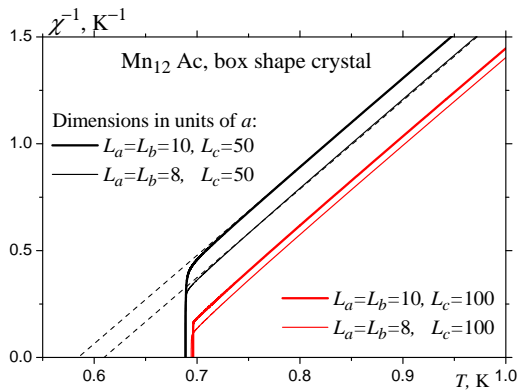,angle=-90,width=8cm}
\end{picture}
\caption{Inverse susceptibility of elongated box-shape crystals of Mn$_{12}$
Ac vs temperature. While the $\protect\chi ^{-1}(T)$ curves extrapolate to
the would-be ferromagnetic ordering temperatures, the real ordering occurs
at a higher temperature and is non-ferromagnetic. }
\label{Fig-Sus}
\end{figure}

Numerical calculations have been done using Wolfram Mathematica for
box-shape crystals of dimensions $L_{a}=L_{b}\ll L_{c}$ in units of the
lattice spacing $a.$ The numbers $L_{a}=L_{b}\ $and $L_{c}$ have been taken
even. The indices $i_{a}$ and $i_{b}$ specifying positions of molecules
within the $a,b$ plane run in the range $-i_{a,b,\max }\leq $ $i_{a,b}\leq
i_{a,b,\max }$ for sublattice 1 and in the range $-i_{a,b,\max }+1/2\leq $ $%
i_{a,b}\leq i_{a,b,\max }-1/2$ for the body-centered sublattice 2, where $%
i_{a,b,\max }=L_{a}/2.$ For the crystallographic direction $c,$ the index
ranges are $-i_{c,\max }\leq $ $i_{c}\leq i_{c,\max }$ and $-i_{c,\max
}+1/2\leq $ $i_{c}\leq i_{c,\max }-1/2,$ respectively, where $i_{c,\max }=%
\mathrm{Round}\left[ L_{a}/\left( 2\eta \right) \right] $ and $\eta
=c/a=0.7153$ for Mn$_{12}$ Ac. The total number of molecules in the crystals
studied thus is given by $N=\left( 2i_{a,\max }+1\right) $ $\left(
2i_{b,\max }+1\right) $ $\left( 2i_{c,\max }+1\right) +\left( 2i_{a,\max
}\right) $ $\left( 2i_{b,\max }\right) $ $\left( 2i_{c,\max }\right) $ that
reached $N=31061$ for the biggest crystal with $L_{a}=L_{b}=10\ $and $%
L_{c}=100.$ However, the actial number of lattice sites in the calculations
was reduces by a factor of approximately 8 by using symmetry that resulted
in the gain of approximately 8$^{2}=64$ in the computer time and memory
usage.

Numerical calculations show that the uniform-field coupling coefficient $%
\left( \mathbf{A}_{\mu }^{L}\cdot \mathbf{I}\right) \left( \mathbf{A}_{\mu
}^{R}\cdot \mathbf{I}\right) $ reaches a large value for some $\mu =\mu _{%
\mathrm{ferro}},$ while for other $\mu $ values $\left( \mathbf{A}_{\mu
}^{L}\cdot \mathbf{I}\right) \left( \mathbf{A}_{\mu }^{R}\cdot \mathbf{I}%
\right) $ are much smaller. The ferromagnetic eigenvectors $\mathbf{A}_{\mu
_{\mathrm{ferro}}}^{L,R}\equiv \mathbf{A}_{\mathrm{ferro}}^{L,R}$ have all
their elements of the same sign, while other eigenvectors have elements of
different signs and do not project well on the uniform field. Thus not too
close to $T_{C}$ Eq.\ (\ref{chiRes}) is dominated by the single
ferromagnetic term,
\begin{equation}
\chi \cong \frac{1}{N}\frac{\left( \mathbf{A}_{\mathrm{ferro}}^{L}\cdot
\mathbf{I}\right) \left( \mathbf{A}_{\mathrm{ferro}}^{R}\cdot \mathbf{I}%
\right) }{T-T_{\mathrm{ferro}}}
\end{equation}
and $\chi ^{-1}(T)\varpropto T-T_{\mathrm{ferro}}$ is a straight line that
extrapolates to $T_{\mathrm{ferro}}$ as an apparent transition temperature.
In all cases for which numerical calculations have been performed, it turns
out that $T_{\mathrm{ferro}}<T_{C}$ and thus $T_{C}$ corresponds to other
types of ordering than ferromagnetic. At $T$ approaches $T_{C},$ the term
with $T_{\mu }=T_{C}$ in Eq.\ (\ref{chiRes}) becomes dominant and the curve $%
\chi ^{-1}(T)$ drops suddenly to zero. The ordering eigenvector
corresponding to the ordering at $T_{C}$ was shown to be non-ferromagnetic.
Of course, the behavior shown in Fig.\ \ref{Fig-Sus} for several different
crystals cannot be seen and actual $T_{C}$ cannot be found, if only the
high-temperature susceptibility data is available, as in Ref.\ %
\onlinecite{bowetal09-arXiv}. In fact, there are many different ordering
eigenvalues $T_{\mu }$ in the region between $T_{\mathrm{ferro}}$ and $T_{C}$

It appears that with increasing the crystal size coupling of the uniform
field to non-ferromagnetic eigenvectors decreases, so that the drop of $\chi
^{-1}(T)$ at $T_{C}$ becomes sharper$.$ This makes using uniform
susceptibility to detect $T_{C}$ in the case of a non-ferromagnetic ordering
questionable. Of course, theoretically one can couple to the
non-ferromagmetic ordering modes by a non-uniform magnetic field (such as
the staggered field in the case of antiferromagnetism) but practically it is
difficult to realize.

One can see from Fig.\ \ref{Fig-Sus} that $T_{\mathrm{ferro}}$ essentially
depends on the aspect ratio, as it should be, and is in accord with Eq.\ (%
\ref{Dzzbar}). To the contrary, $T_{C}$ does not strongly depend on the
shape since it corresponds to a non-ferromagnetic ordering. Its moderate
increase with the crystal size shows that the crystal sizes in the numerical
calculations are still somewhat small to perfectly reproduce the behavior of
macroscopic crystals. With increasing elongation, ferromagnetic ordering
becomes more competitive but still falls behind other orderings up to the
aspect ratios of about 12 where ellipsoids would already order
ferromagnetically. It is difficult to increase the elongation while keeping
the transverse size large enough in the calculations because the number of
molecules becomes too large.

\begin{figure}[t]
\unitlength1cm
\begin{picture}(11,5)
\psfig{file=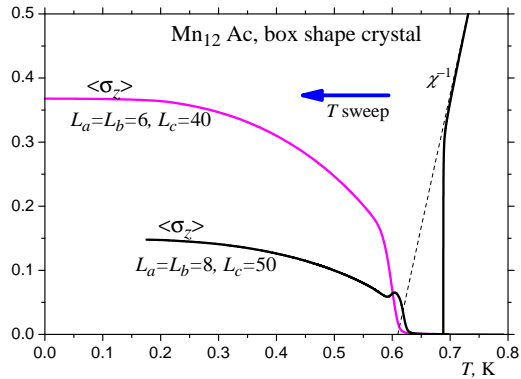,angle=-90,width=8cm}
\end{picture}
\caption{Magnetization of elongated box-shape crystals of Mn$_{12}$ Ac of
different dimensions, developing upon lowering temperature at a slow
constant rate. Static inverse susceptibility for $L_{a}=L_{b}=8,$ $L_{c}=50$
from the preceding figure is also shown.}
\label{Fig-TLowering}
\end{figure}
\begin{figure}[t]
\unitlength1cm
\begin{picture}(11,5)
\psfig{file=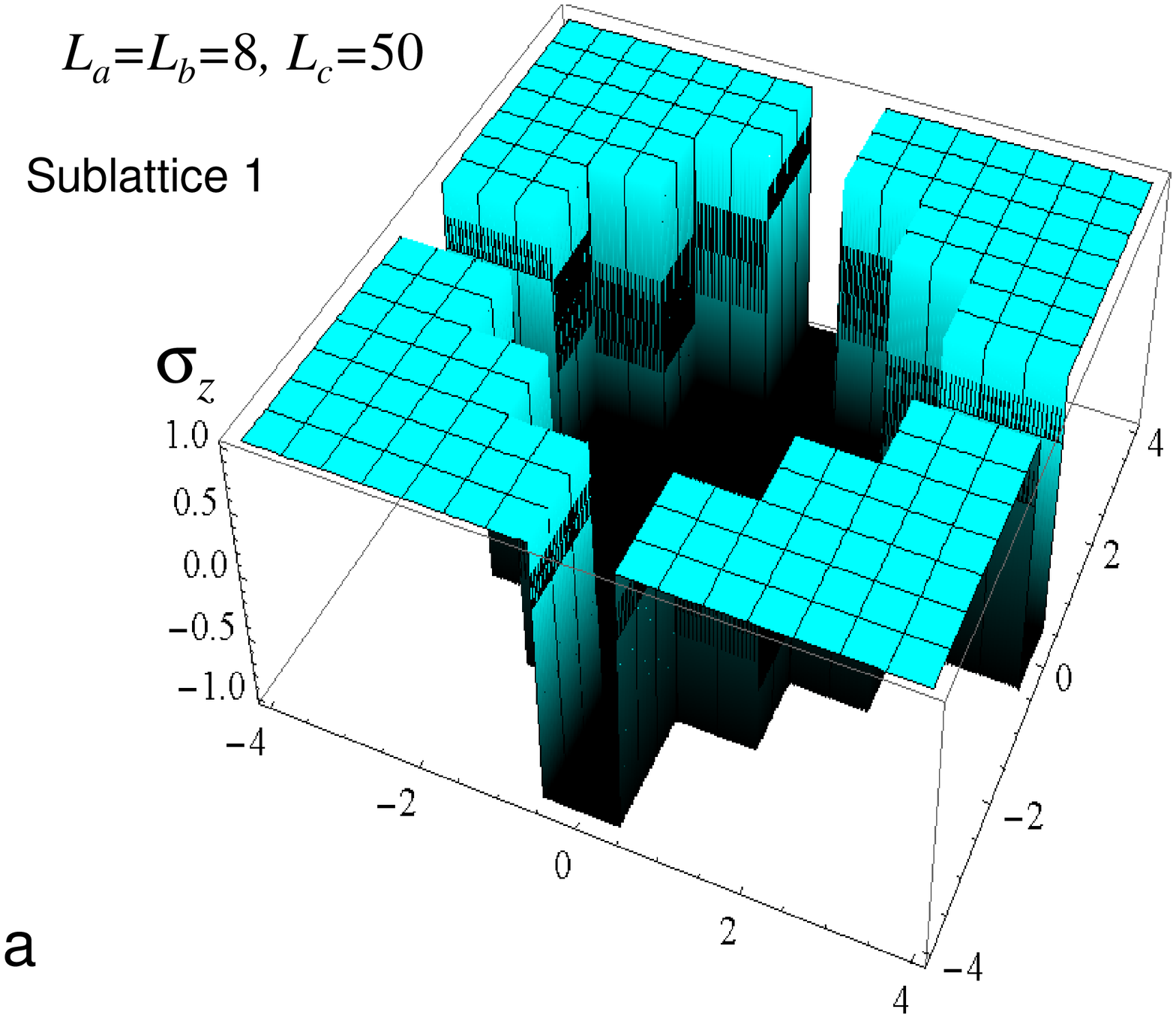,angle=-90,width=8cm}
\end{picture}
\unitlength1cm
\begin{picture}(11,5)
\psfig{file=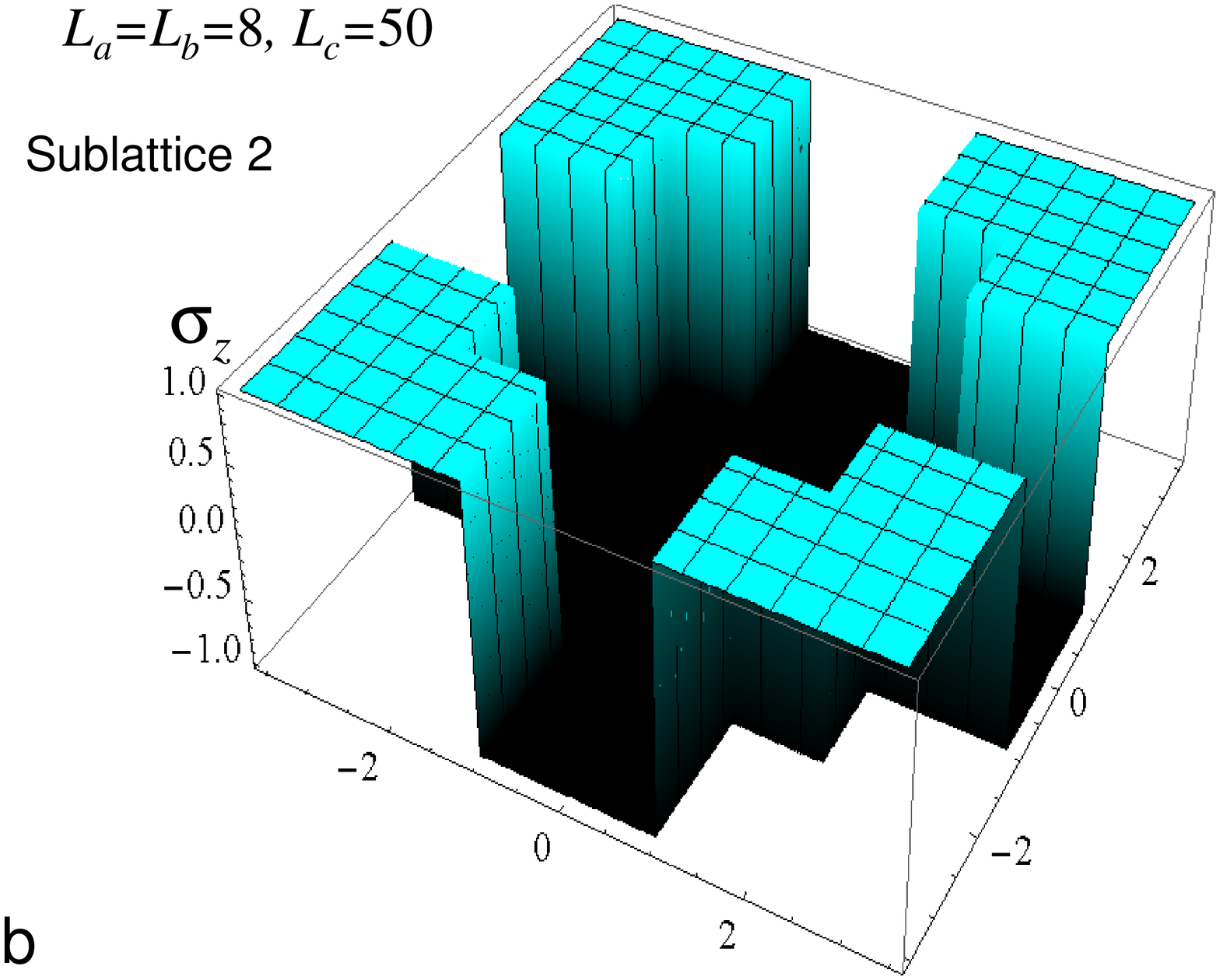,angle=-90,width=8cm}
\end{picture}
\caption{Ordering in elongated box shape Mn$_{12}$ Ac crystals ($%
L_{a}=L_{b}=8,$ $L_{c}=50)$ at $T=0,$ obtained by slow lowering temperature
from $T_{\max }>T_{C}$ and shown in the cross-section through the middle of
the crystal. a) Sublattice 1; b) Sublattice 2. Magnetization in both
sublattices is qualitatively similar. Central regions are magnetized
opposite to the small external field while periferal regions are magnetized
along it. Magnetization at $T=0$ is uniform along the $c$ direction.}
\label{Fig-Ordering}
\end{figure}

Note that in the case of the standard antiferromagnetism the ferromagnetic
eigenvalue $T_{\mathrm{ferro}}$ is negative. The ferromagnetic state in the
antiferromagnet is absolutely unstable, as the molecular field is opposite
to the spins. In our problem of dipolar ordering, there are many positive
and many negative eigenvalues $T_{\mu }.$ Negative $T_{\mu }$ correspond to
absolutely unstable states, whereas $T_{\mu }>0$ correspond to local minima
of energy with dipolar fields parallel to spins.

Competition of many local energy minima in our model of ordering in Mn$_{12}$
Ac makes it impossible to describe the ordered state below $T_{C}$ by
solving Eq.\ (\ref{CWEq}) directly. Instead of the global minimum of the
free energy, the solver finds local minima, local maxima, or saddle points.
A more reliable method is to solve the system of relaxational equations
\begin{equation}
d\left\langle \sigma _{iz}\right\rangle /dt=-\Gamma \left[ \left\langle
\sigma _{iz}\right\rangle -\tanh \left( \ldots \right) \right] ,
\label{DynamicsEq}
\end{equation}
where $\Gamma $ is the relaxation rate and $\tanh \left( \ldots \right) $ is
the same as in Eq.\ (\ref{CWEq}). As the purpose of this work was to study
ordering rather than the exact dynamics, $\Gamma $ was set to an arbitrary
constant in numerical calculations. It was found that Eq.\ (\ref{DynamicsEq}%
) leads to freezing into a spin-glass state as the result of relaxation out
of a random initial state at low temperatures. To obtain the ordering type
that is mostly close to the lowest free energy state at any temperature, one
can solve Eq.\ (\ref{DynamicsEq}) with temperature $T$ slowly changing in
time from some $T_{\max }>T_{C}$ to nearly zero. Just to initiate ordering,
one can set $h_{z}$ to a very small value.

Fig.\ \ref{Fig-TLowering} shows the results of these calculations for two
different crystal sizes. The $\left\langle \sigma _{z}\right\rangle $ curves
accurately reproduce the equilibrium magnetization everywhere except for the
critical region (0.6 K $\leq T\leq 0.7$ K) where critical slowing down
requires a slower temperature sweep that is difficult to implement in the
numeric routine. The average magnetization at $T\rightarrow 0$ in both cases
is significantly smaller than 1 because of a non-ferromagnetic type of
ordering. A nonzero average magnetization in the ordered state seems to be a
finite-size effect related to the incomplete compensation of molecules with
spins up and down in small crystals. Ordering attained at $T\rightarrow 0$
is shown in Fig.\ \ref{Fig-Ordering} for cross-sections through the middle
of the crystal length. In fact, the magnetization at $T\rightarrow 0$ is
uniform along the $c$ axis. The inner and outer regions of the crystal order
in different directions, in a similar way in both sublattices.

Concluding, mean-field calculations do not support ferromagnetic ordering in
elongated box-shape crystals of Mn$_{12}$ Ac that are currently under
investigation.\cite{bowetal09-arXiv} How long must be box-shape crystals to
order ferromagnetically remains an open question. In any case, ferromagnetic
ordering should be extremely difficult to observe in experiments because of
a strong competition from numerous other types of ordering.

Useful discussions with E. M. Chudnovsky, A. D. Kent, M. P. Sarachik, Y.
Yeshurun, and A. J. Millis are greatfully acknowledged. I have profited from
having access to the experimental data by A. D. Kent, M. P. Sarachik, Y.
Yeshurun, Bo Wen, and other team members during the experiments.

This work has been supported by the NSF Grant No. DMR-0703639.

\bibliographystyle{apsrev}
\bibliography{chu-own,gar-own,gar-MM-ordering,gar-tunneling}

\end{document}